\documentclass[journal]{IEEEtran}

\usepackage{setspace,amsmath,latexsym,cite,amssymb,epsfig,amsfonts}
\usepackage{url,cite}
\usepackage{graphicx}
\usepackage{psfrag}
\usepackage{footmisc}
\usepackage{multirow}
\usepackage{color}

\graphicspath{{.}{../eps/}{./images/}{./images/result/}}
\usepackage{amssymb}
\usepackage{amsthm}

\usepackage{cite}
\usepackage{mathtools, cuted}
\usepackage[font=small, labelsep=period]{caption}
\usepackage{makecell}
\usepackage{stfloats}
\usepackage{subcaption}
\usepackage{changepage}
\usepackage{balance}
\allowdisplaybreaks[4]

\author{Chong Huang, \IEEEmembership{Member, IEEE}, Yun Wen, \IEEEmembership{Graduate Student Member, IEEE}, Long Zhang,\\Gaojie Chen, \IEEEmembership{Senior Member, IEEE}, Zhen Gao, Pei Xiao, \IEEEmembership{Senior Member, IEEE}
\thanks{C. Huang, Y. Wen, G. Chen and P. Xiao are with 5GIC \& 6GIC, Institute for Communication Systems (ICS), University of Surrey, Guildford, GU2 7XH, United Kingdom, Email: \{chong.huang, yun.wen, gaojie.chen, p.xiao\}@surrey.ac.uk.}
\thanks{L. Zhang is with the PengCheng Laboratory, Shenzhen, Guangdong 518066, China, E-mail: zhangl02@pcl.ac.cn.}
\thanks{Z. Gao is with the MIIT Key Laboratory of Complex-Field Intelligent Sensing, Beijing Institute of Technology, Beijing 100081, China, also with the Yangtze Delta Region Academy, Beijing Institute of Technology (Jiaxing), Jiaxing 314019, China, and also with the Advanced Technology Research Institute, Beijing Institute of Technology, Jinan 250307, China, E-mail: gaozhen16@bit.edu.cn.}
}

\begin{document}

\title{\Huge Reconfigurable Intelligent Surface Empowered Full Duplex Systems: Opportunities and Challenges}

\maketitle
{
\begin{abstract}
Reconfigurable intelligent surfaces (RISs) have emerged as a promising technology in wireless communications. Simultaneously transmitting and reflecting RIS (STAR-RISs) in particular have garnered significant attention due to their dual capabilities of simultaneous transmission and reflection, underscoring their potential applications in critical scenarios within the forthcoming sixth-generation (6G) technology landscape. Moreover, full-duplex (FD) systems have emerged as a breakthrough research direction in wireless transmission technology due to their high spectral efficiency. This paper explores the application potential of STAR-RIS in FD systems for future wireless communications, presenting an innovative technology that provides robust self-interference cancellation (SIC) capabilities for FD systems. We utilize the refraction functionality of STAR-RIS enhances the transmission capacity of FD systems, while its reflection functionality is used to eliminate self interference within the FD system. We delve into the applications of two different types of STAR-RIS in FD systems and compare their performance through simulations. Furthermore, we discuss the performance differences of STAR-RIS empowered FD systems under various configurations in a case study, and demonstrate the superiority of the proposed deep learning-based optimization algorithm. Finally, we discuss possible future research directions for STAR-RIS empowered FD systems.
\end{abstract}
}
\IEEEpeerreviewmaketitle
\section{Introduction}\label{sec:1}
Over the past few decades, a rapid growth in the number of wireless devices has created an ever-increasing need for uninterrupted widespread connectivity and high-speed data transmission. To accommodate the future demands of diverse applications, next-generation broadband networks must continually enhance their spectral efficiency (SE), reduce latency, and lower infrastructure costs and energy consumption. Especially, the scarcity of available frequency bands has compelled researchers to focus on enhancing SE by optimizing the utilization of the limited spectrum resources \cite{SE_refer1}.

Among numerous approaches to enhance SE, in-band full-duplex (FD) technology has gathered significant interests from both academic and industries in recent years \cite{FD_survey2}. Currently deployed wireless systems typically utilize either a time-division duplex (TDD) or frequency-division duplex (FDD) approach for bidirectional communication. This half-duplex (HD) nature of wireless system involves partitioning temporal and/or spectral resources into orthogonal segments and allocating them to bidirectional communication user devices (UE). In contrast, FD communications refers to the transmission and reception of a certain wireless device operated simultaneously in the same frequency band. Besides the possible advantages of doubling the SE under heavy traffic, the unique characteristic of the FD communication can also provide novel functions such as sensing during the transmission or jamming during the reception for different purposes, with its performance widely investigated in numerous scenarios \cite{FD_survey_multisce}.

Despite its remarkable performance, FD must confront the inherent challenge of self interference. The self interference introduced by an FD terminal's transmission significantly impacts its own signal reception, thus necessitating the implementation of effective cancellation techniques. Numerous self interference cancellation (SIC) techniques have been developed, which can provide a SIC capacity of suppressing the interference power and reduced its power to the noise floor for some applications such as the femto-cell \cite{SIC_new1}. However, due to their high hardware cost by introducing extra circuit between antennas and the limited SIC capacity, their application are limited to those with small number of antennas and with low transmit power, which can not satisfy the request of extreme high data rate and seamless coverage of the next generation networks. Recent meeting records from 3GPP related to SIC indicate that the residual self interference in FD systems remains an issue that requires further research in exiting wireless devices.

As an intriguing wireless communication solution to provide extra coverage while maintaining low complexity and power consumption, the reconfigurable intelligent surface (RIS) has emerged significant interest in the last few years \cite{RIS_survey}. Comprising an extensive array of passive reflecting components whose amplitudes and phases are controlled by software, RIS offers a cost-effective means to establish a programmable propagation environment, thereby improving the efficiency of wireless communication. With its substantial potential, RIS-assisted communication is increasingly recognized as a promising enabling technology in future sixth-generation (6G) communications \cite{9475160}.

Due to the reflecting nature of the RIS, a major shortage that limits its application is that it can only serve devices located on one side. In recent developments, a novel concept known as simultaneously transmitting and reflecting (STAR)-RIS has emerged as an advancement over traditional RIS technology \cite{STAR-2}. STAR-RIS elements possess dual functionalities of transmitting and reflecting, allowing them to simultaneously reflect and refract incident signals. This innovative capability enables STAR-RIS to efficiently serve devices on both sides of the surface, making it significantly more effective in enhancing radio coverage compared to conventional RISs, as discussed in \cite{STAR-1}.

Without involving additional radio-frequency (RF) chains and amplifiers in its operation, an RIS works fundamentally as an FD device with negligible antenna noise and self interference \cite{10194555}. Considering its FD nature and the ability to optimize propagation environment, incorporating RIS with FD communication is regarded with the great potential of exploiting both technologies to significantly improve the performance of wireless systems from various aspects such as SE and energy efficiency (EE) \cite{FD_survey2}.

The rest of this article is structured as follows: Section II gives a brief glance of the principle of FD and RIS technologies, as well as a overview of recent works that joint designed RIS and FD communication in various application areas. Section III focuses on reviewing the proposed soft-combination scheme of RIS-assisted FD communication, while a case study of FD communication with RIS-assisted SIC is presented in section IV. In Section V, we explore the remaining challenges and future research trends RIS-assisted FD communication. In the last, we conclude this article in Section VI.

\section{Principles and Challenges}\label{sec:Principle}
\subsection{Principles of RIS and FD Systems}
\subsubsection{FD communications}
\begin{figure*}[t!]
\begin{adjustwidth}{0em}{0em}
\begin{subfigure}[b]{.3\textwidth}
  \centering
  \includegraphics[width=1\linewidth, height=0.7\linewidth]{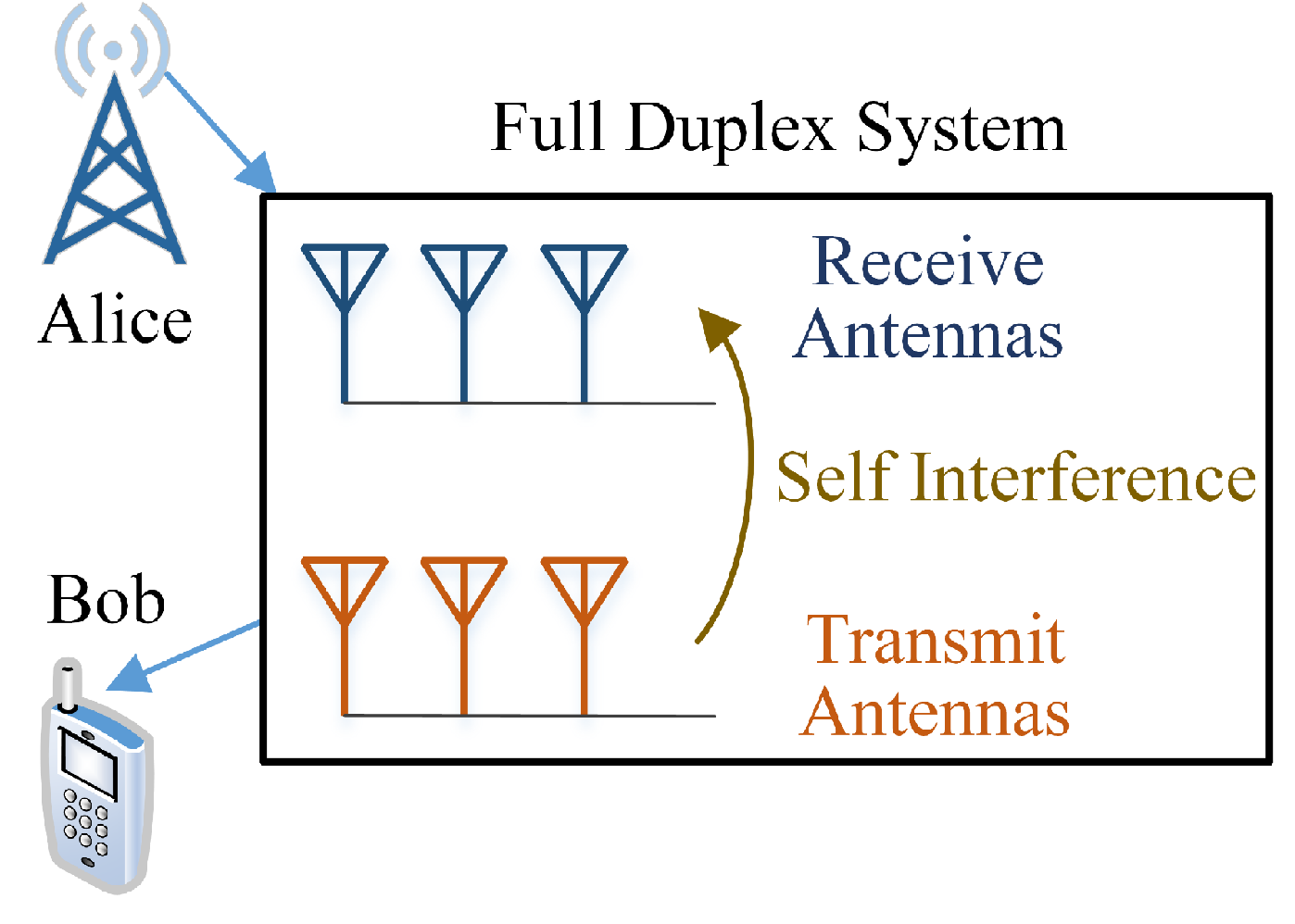}
  \caption{The structure of FD systems. }
  \label{fig:IntroFD}
\end{subfigure}\hspace{10mm}
\begin{subfigure}[b]{.3\textwidth}
  \centering
  \includegraphics[width=1\linewidth, height=0.7\linewidth]{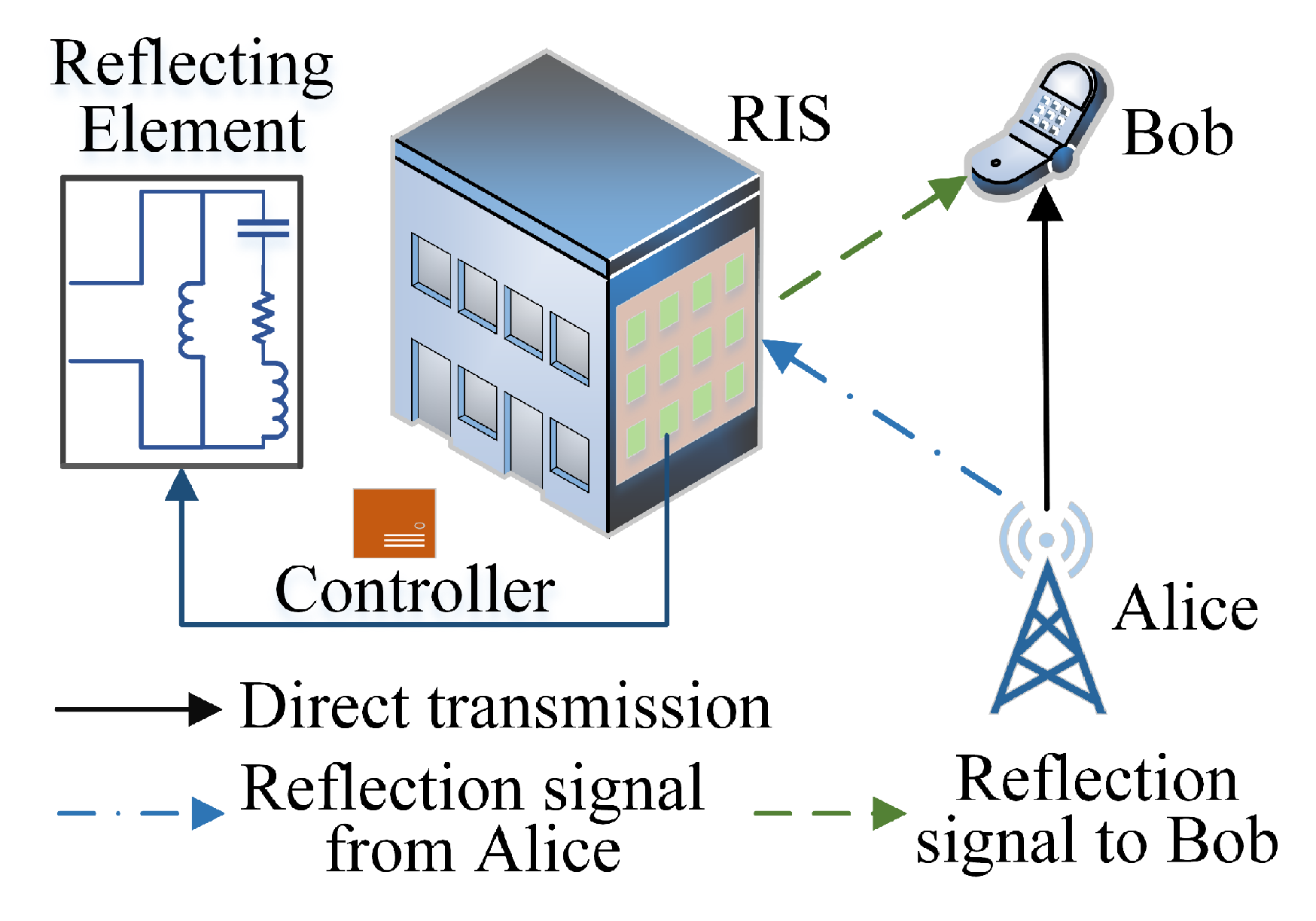}
  \caption{The structure of RIS systems.}
  \label{fig:IntroRIS}
\end{subfigure}\hspace{10mm}
\begin{subfigure}[b]{.3\textwidth}
  \centering
  \includegraphics[width=1\linewidth, height=0.7\linewidth]{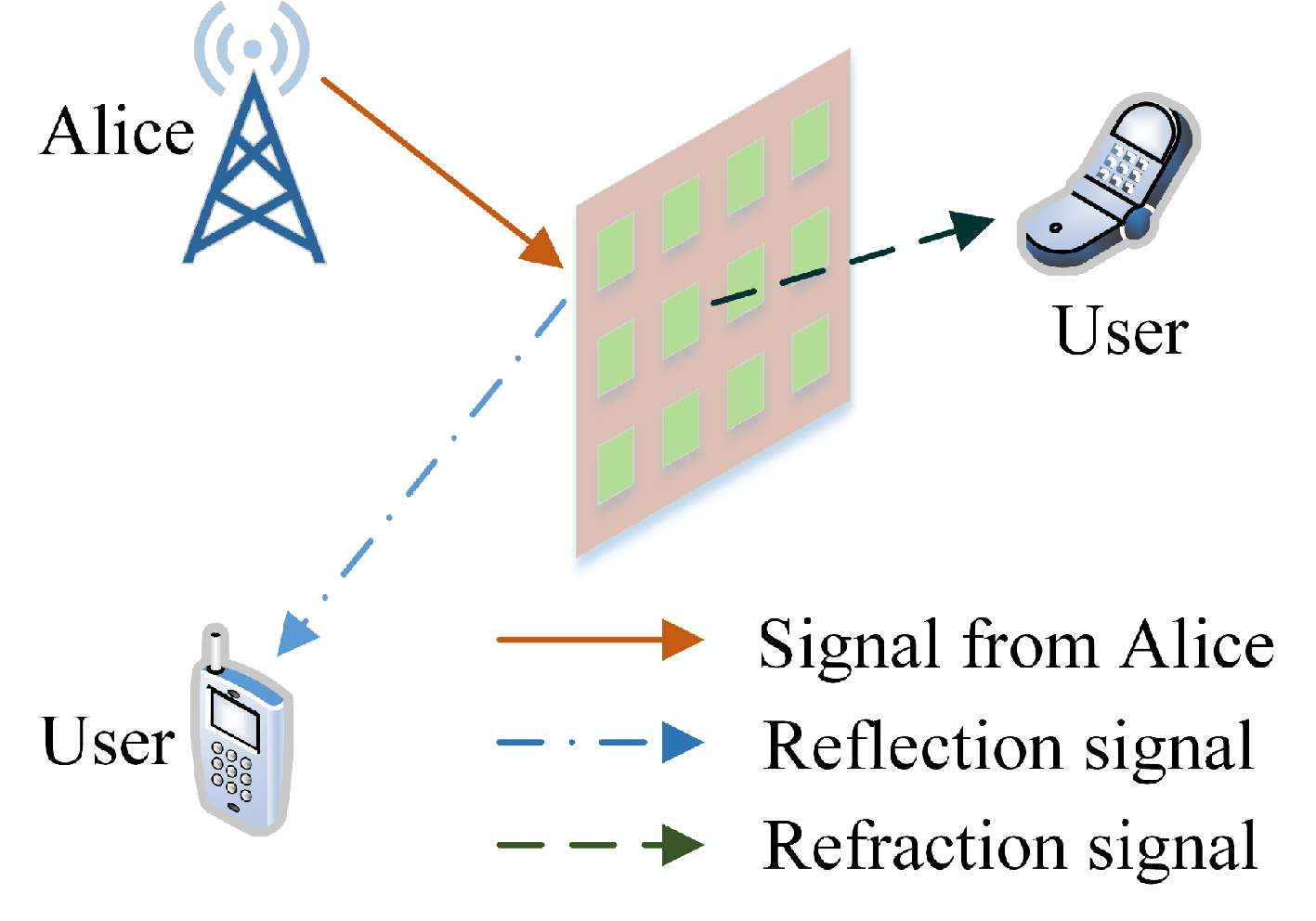}
  \caption{The structure of STAR-RIS systems.}
  \label{fig:IntroStarRIS}
\end{subfigure}
\end{adjustwidth}
\caption{The structure of FD, RIS and STAR-RIS systems.}
\label{fig:structures}
\end{figure*}

FD wireless communication represents a transformative advancement in telecommunication technologies, allowing simultaneous two-way data transmission over the same frequency channel. This capability not only enhances SE but also significantly improves communication latency and throughput. The structure of an FD system is shown in Fig. \ref{fig:IntroFD}, the signal transmitted by transmit antennas can cause interference to the receive antennas. Thus, central to achieving FD operation is the effective cancellation of self interference, the signal that a transmitter generates which can overpower the signal it simultaneously receives.

The framework for SIC in FD communications can be divided into three main strategies:
i) Passive Suppression: Initial mitigation of self interference involves passive techniques that inherently reduce the leakage of the transmitter signal into the receiver. This includes the design and placement of antennas to maximize spatial isolation between transmit and receive paths. Techniques such as using directional antennas, physically separating transmit and receive antennas, and employing cross-polarization can significantly attenuate the self interference before any active cancellation is applied.
ii) Analog-domain cancellation: Analog-domain cancellation involves creating a signal that is identical in magnitude but opposite in phase to the self interference signal. This inverse signal is then injected into the receiver pathway where it ideally couples the self interference, effectively neutralizing its impact on the received signal. Precise control over the amplitude and phase of the inverse signal adaptive to the dynamic transmission environment and equipment conditions is essential in this stage, which can reduce the residual self interference to a sufficiently low level for analog-to-digital converter (ADC) for the following digital-domain cancellation.
iii) Digital-domain cancellation: Considerable residual interference may still exist after passive suppression and analog-domain cancellation. Therefore, advanced digital signal processing (DSP) techniques are employed to detect, analyze, and filter out this remaining interference from the received signal. The digital-domain cancellation first estimates the self interference channel, and then applies the estimated channel to the known transmitted signals to generate digital signals, thus remove the self interference in the received signals.

\subsubsection{RIS and STAR-RIS}
An RIS is a sophisticated, digitally-manipulable metasurface composed of an extensive array of passive reflective elements, as shown in Fig. \ref{fig:IntroRIS}. Behind each element, there's typically a small electronic circuit. These components can alter their electrical characteristics, i.e. capacitance or resistance, when controlled by an external voltage. By adjusting these characteristics, the RIS can dynamically modify the impedance of each element with ultra-low power, thus introduce phase shifts and/or amplitudes of the incoming signals.

Distinct from conventional wireless communication methodologies implemented at transceivers, an RIS possesses the unique capability to directly alter the wireless propagation channel, thereby enhancing the efficacy of signal transmission. This includes amplifying the signal power received by designated users and attenuating interference experienced by non-intended recipients.

Due to its reflection nature, the users located on the opposite side of the RIS are out of coverage. To address this issue, STAR-RIS, also referred as intelligent omni surface (IOS), has recently been developed as an enhancement to traditional RISs. An latest prototype of STAR-RIS implement elements comprise multiple metallic patches and PIN diodes, which can dynamically adjust the amplitude and the phase shifts for both reflecting and refracting signals \cite{IOS_prototype}. This advantage makes it possible to serve devices on both sides of the surface, as shown in Fig. \ref{fig:IntroStarRIS}, thus being significantly more effective in enhancing radio coverage compared to conventional RISs. The STAR-RIS is generally studied with three operation modes: 1) energy splitting (ES) mode where each element can reflect and refract simultaneously; 2) mode switching (MS) mode where each element can only choose to reflect or refract in a certain time period; 3) time switching (TS) mode where all elements choose to reflect or refract in a certain time period \cite{STAR-1}. This innovative capability enables STAR-RIS to efficiently serve devices on both sides of the surface.

\subsection{Challenges of Traditional FD Systems}
Despite the their promising potentials, both RIS and FD communication face numerous open challenges as pointed out by the latest analysis, among which the most significant ones are the practical limitations of traditional SIC schemes.

\subsubsection{Hardware limitation and impairments}
First of all, the passive suppression stage of the SIC usually requires a considerable antennas spacing, e.g. 20 cm to 40 cm between the transmit and the receive antenna, which becomes impractical for most of the mobile devices. Many passive suppression techniques, such as directional passive suppression, depend extensively on configurations with multiple antennas, thereby rendering receivers with size constraints unable to adequately mitigate self interference power. After that, the inherent compromise between hardware expenditures and SI efficacy persists as a significant practical challenge for analog cancellation. Specifically, enhancing cancellation capabilities necessitates stringent requirements on hardware precision. Certain analog cancellation methodologies, including those based on delay lines, demand elevated delay resolution accuracy. Nonetheless, achieving this precision is contingent upon the utilization of more complex and extensive hardware circuitry. For effective digital cancellation, an adequate dynamic range and resolution ADC are essential to accurately capture both the SI and the desired signals. This is necessary to facilitate the cancellation of the transmitted signal that is injected into the receiver's baseband. In practical scenarios, the extent of achievable digital cancellation is influenced by various hardware impairments and constraints. These include nonlinearities, the limited dynamic range and resolution of ADC circuits, phase noise, and multi-path environmental reflections. Due to the circuit complexity and the hardware impairments, the capacity of even the most recent SIC technologies is limited as summarized in \cite{FD_survey1}, which prevents the FD communication from being applied to devices with high transmit power such as base stations.

\subsubsection{Power consumption}
The high power consumption and insufficient capacity of traditional SIC technologies are also obstacles of its implementation. By introducing extra circuits between the transmit and receive antennas, traditional SIC schemes can couple the transmitted signals into the received signals with delay and phase compensation as well as power attenuation, to cancel the self interference from the propagation domain. However, the extra circuit components not only introduce a considerable increase in chip area but also a significant growth of power consumption. Numerous latest implements of SIC for FD communication have been proposed and compared in \cite{SIC_power_3}, which shows that the power consumption power of SIC increases in proportion to the number of transmit/receive antenna pairs and may become even higher than that for transmission in certain conditions.
\subsubsection{Computational Complexity}
SIC technology plays a key role in the implementation of FD wireless systems, where the computational complexity of FD transmission is largely manifested in the elimination of self interference. In FD communications, SIC requires high speed and high precision processing of the received signal to identify and remove interference caused by the local transmit antennas in real time. This process typically involves multi-stage signal estimation and multi-variable optimization algorithms, which lead to high computational complexity \cite{7901493}, especially in a large-scale network where computational demands grow exponentially. As wireless communication technology evolves, particularly with the upcoming 6G and higher standards wireless networks, there is a need for further optimization of these algorithms to accommodate higher data rates and lower latency requirements. Therefore, the implementation of FD systems in future wireless network architectures will continue to face the dual challenges of optimizing computational resources and enhancing system efficiency.

\section{STAR-RIS Empowered Full-Duplex System}\label{sec:Proposed}
FD technology allows wireless devices to simultaneously transmit and receive on the same frequency band. However, a major challenge of this mode is self-interference, where the signal from the transmit antenna interferes with the receive antenna. Traditional SIC techniques include using special hardware architectures and signal processing algorithms to reduce interference, but these methods are often complex and costly.

STAR-RIS technology addresses this issue through its dual functionalities of transmitting and reflecting. STAR-RIS is composed of numerous small elements, each can independently adjust its reflective and transmissive properties to reconfig the transmission environment. In FD communication scenarios, STAR-RIS can be precisely configured on the FD device to eliminate the interference from the transmit antenna to the receive antenna through the reflective pathway, while using the the transmissive pathway to send the transmit signal to the receiver. This approach not only reduces hardware complexity but also operates in a low power-consumption manner.

\begin{figure}[t!]
        \centering
        \includegraphics*[width=80mm]{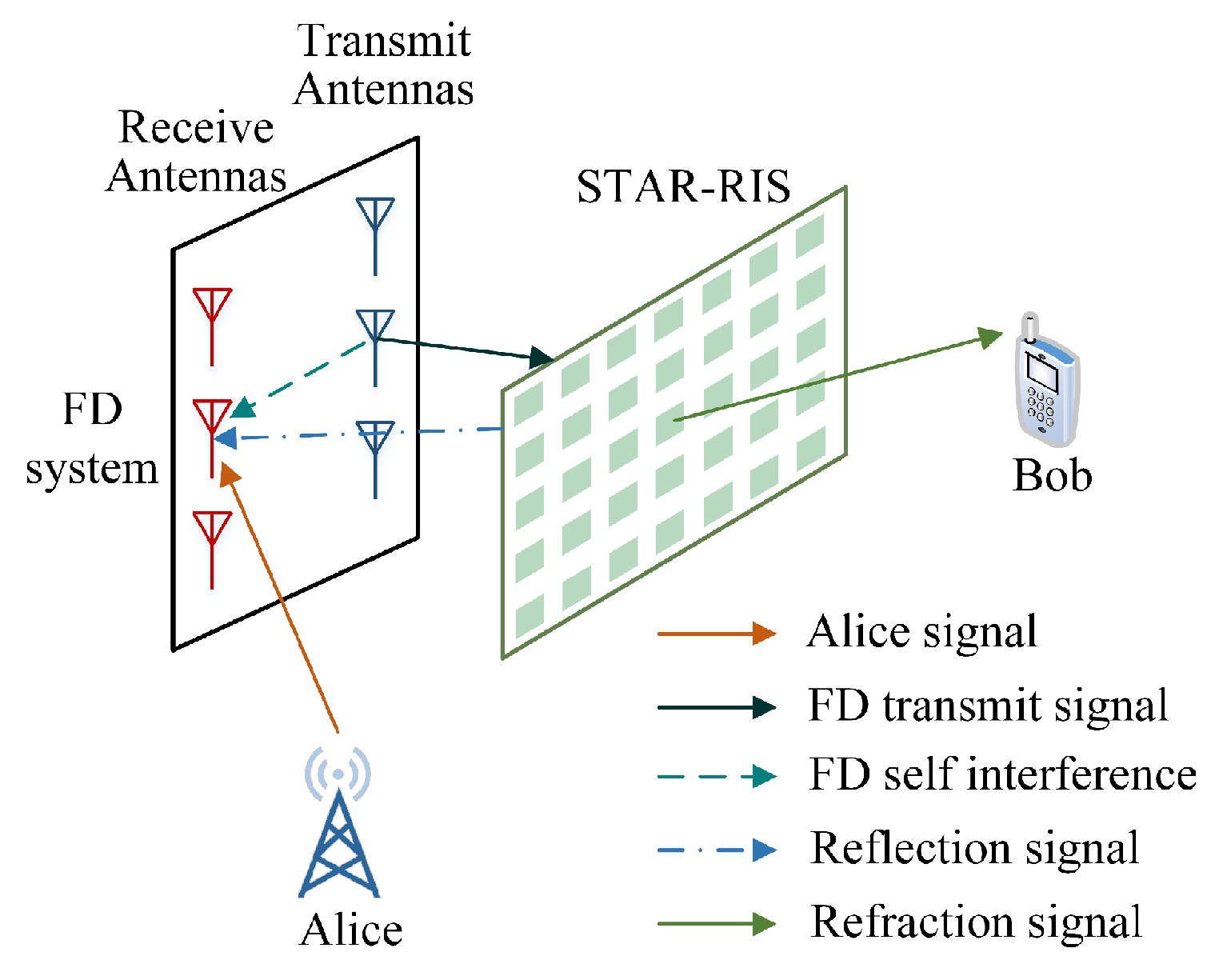}
       \caption{ES-RIS in FD systems.}
        \label{figES}
         \vspace{-1em}
\end{figure}

\begin{figure}[t!]
        \centering
        \includegraphics*[width=80mm]{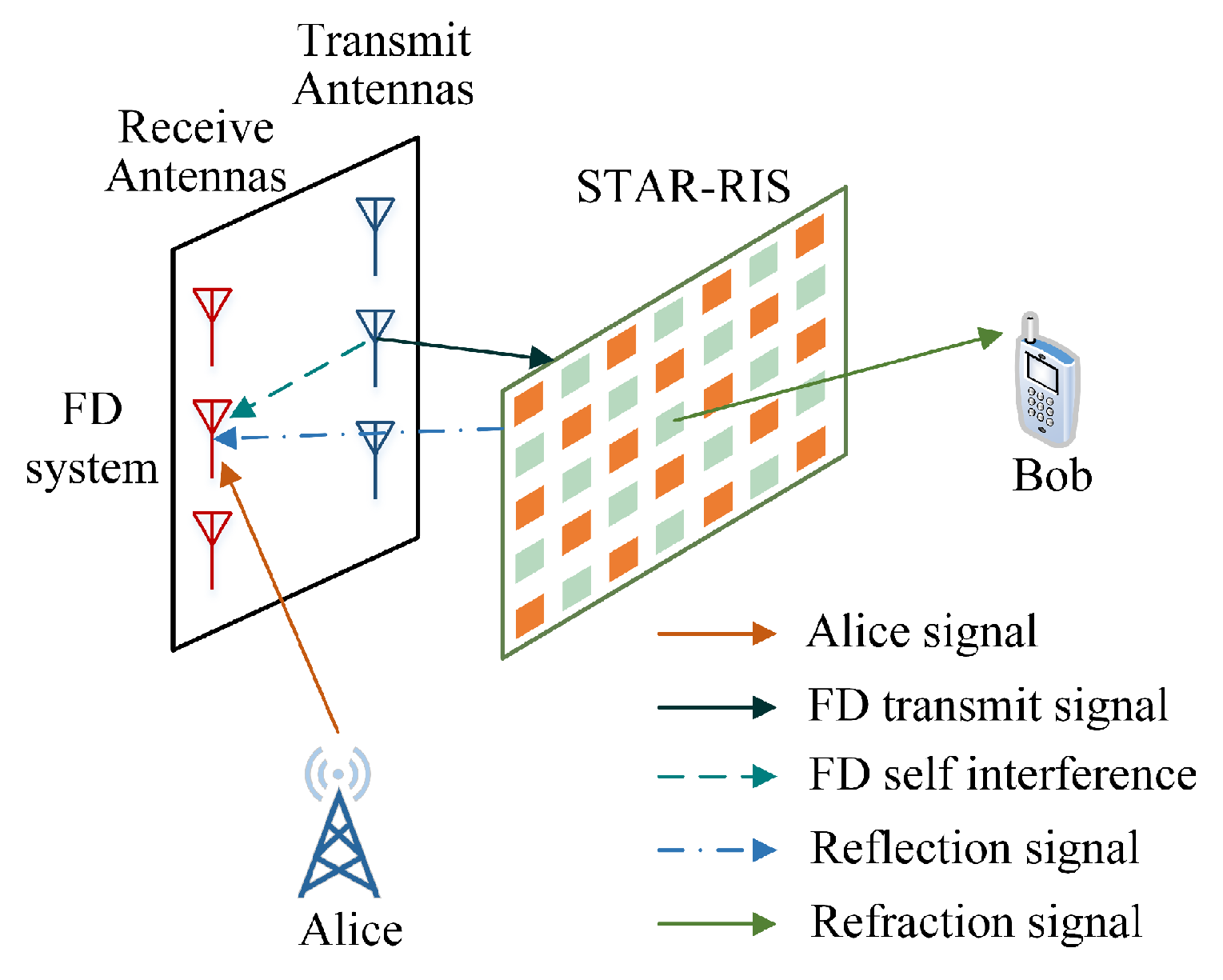}
       \caption{MS-RIS in FD systems.}
        \label{figMS}
         \vspace{-1em}
\end{figure}

Given the importance of RIS-assisted FD communications and the limitation of current works, we considered a combined scheme, which realize SIC of FD communication through the assistance of STAR-RIS, as a promising approach to further exploit the potential of RIS-assisted FD communications. We propose two different STAR-RIS systems in FD communications, ES-RIS and mode switching MS-RIS \cite{10466589}, as show in Fig. \ref{figES} and Fig. \ref{figMS}. We consider the node Alice in Fig. \ref{figES} and Fig. \ref{figMS} transmitting a signal to the FD device which simultaneously transmits a signal to Bob.


From Fig. \ref{figES}, it can be observed that all elements can simultaneously reflect and transmit signals in ES-RIS. In FD systems, the signal emitted by the transmit antennas can interfere with the receive antennas through a direct path. To address this interference problem, we introduce the STAR-RIS in FD communications as illustrated in Fig. \ref{figES}, the STAR-RIS provides a reflective path whereby the signal emitted by the transmit antennas is also reflected by the STAR-RIS to the receive antennas. Through proper configuration of the STAR-RIS, the reflected signals can be made to partially cancel out the directly transmitted signals, thus potentially eliminating self-interference at the receive antennas. Furthermore, since the STAR-RIS can also transmit signals while reflecting, the transmit antennas can send signals to the Bob through the STAR-RIS's refraction function. Therefore, STAR-RIS can simultaneously achieve data transmission and SIC functions, making it highly compatible with FD systems.

Fig. \ref{figMS} shows another application of STAR-RIS with MS mode in FD systems. In an MS-RIS, each element can only perform either reflection or refraction at any given time, thus some elements are dedicated to reflecting signals while others handle refracting. Similar to ES-RIS, the STAR-RIS in an FD system uses reflected signals to eliminate self-interference while transmitting the signals emitted by the transmit antennas to the Bob. Unlike ES-RIS, the part-refractive and part-reflective nature of MS-RIS reduces the complexity of control, making it a more configurable and deployable system.

Employing STAR-RIS for SIC offers numerous advantages that are transformative for FD communications. The versatility of STAR-RIS empowered FD wireless systems in handling both signal refraction and reflection, this integration capability is particularly advantageous for developing smart cities and future large scale mobile networks. Moreover, this capability of STAR-RIS enables even finer control over the electromagnetic properties of the signals, thus opening up possibilities for even more sophisticated interference management and signal optimization strategies in complex FD scenarios.

\section{Case Study}\label{sec:Case}
\begin{figure}[t!]
        \centering
        \includegraphics*[width=80mm]{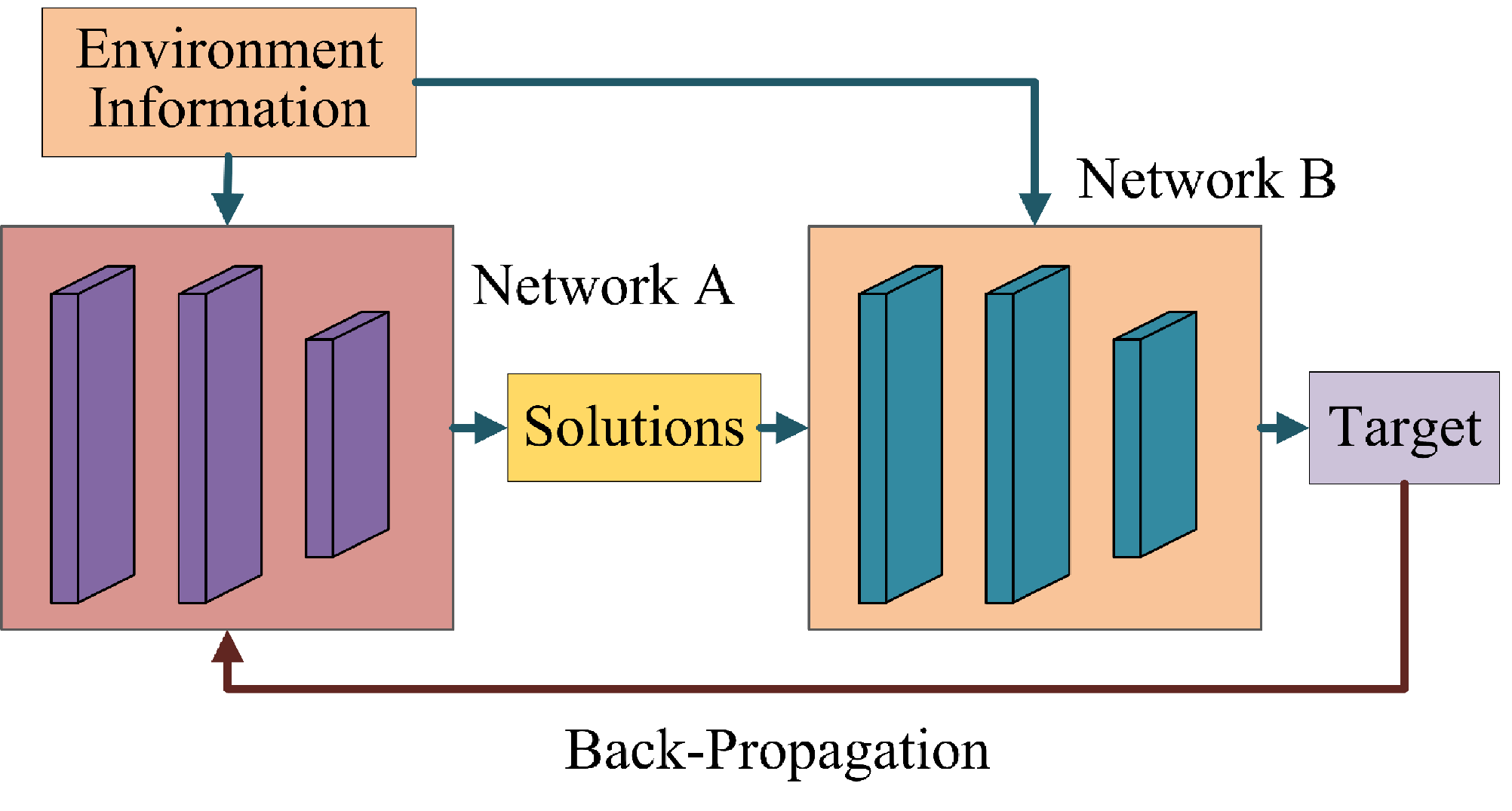}
       \caption{The proposed generative deep learning algorithm for STAR-RIS empowered FD systems.}
        \label{figDL}
         \vspace{-1em}
\end{figure}
Due to the complex electromagnetic environments and dynamism that future wireless communication networks may face, the STAR-RIS integrated with FD systems could encounter highly complex and dynamic conditions, and might be designed to serve multiple users in future large-scale networks. Therefore, traditional algorithms may face difficulties in promptly handling environments with high complexity and dynamic changes. To address this issue, we are applying deep learning techniques to the proposed FD system. Recently, deep learning is primarily applied in wireless communications due to its ability to learn from complex data and predict the behavior of communication systems, which facilitates the automatic optimization of system performance and enhances the speed and reliability of data transmission. As technology progresses, deep learning is expected to be increasingly utilized in intelligent network management and end-to-end communication system design, making future wireless networks more efficient and flexible. To be specific in the proposed FD system, the generative networks, especially in STAR-RIS settings, demonstrate unique advantages such as simulating various communication environments and generating optimized signal configurations to improve system adaptability and performance. This capability makes generative networks an ideal tool for optimizing STAR-RIS configurations, enabling real-time signal optimization and dynamic adjustment to maintain optimal performance in a constantly changing communication environment.

The proposed deep learning algorithm comprises a two-tier network where Network A inputs environmental information of the proposed system, such as channel characteristics, and outputs configuration parameters for the STAR-RIS and beamforming vectors for antennas in the FD system, as shown in Fig. \ref{figDL}. Due to a lack of known samples, it is not feasible to train Network A on a large scale based on optimal solutions as traditional supervised learning does. Hence, we introduced another network, Network B as in \cite{9847234}, which takes the environmental information of the proposed system and the outputs from Network A as the input, even if the performance of these solutions is unstable. Given our explicit objectives, such as maximizing the rate transmitted to the receiver or maximally suppressing self-interference, we can set these objectives as the outputs of Network B. Therefore, after extensive random sampling and training, Network B can model the relationship between system environments, system configurations, and corresponding outcomes. After training Network B, we can fix its parameters and merge Network A with Network B into a single network. By setting the loss function to maximize the rate or the degree of self-interference suppression, and using back-propagation, we update the parameters of Network A to fit the goals maximized by Network B, thus approaching the optimal network environment configuration.

\begin{figure}[t!]
        \centering
        \includegraphics*[width=80mm]{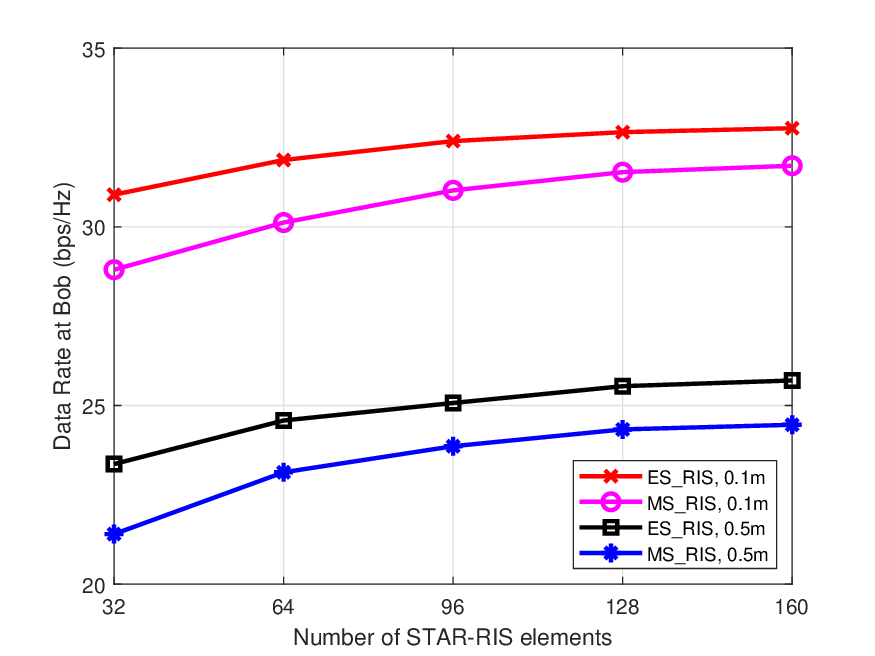}
       \caption{Received data rate at Bob with different number of STAR-RIS elements.}
        \label{figR1}
         \vspace{-1em}
\end{figure}
\begin{figure}[t!]
        \centering
        \includegraphics*[width=80mm]{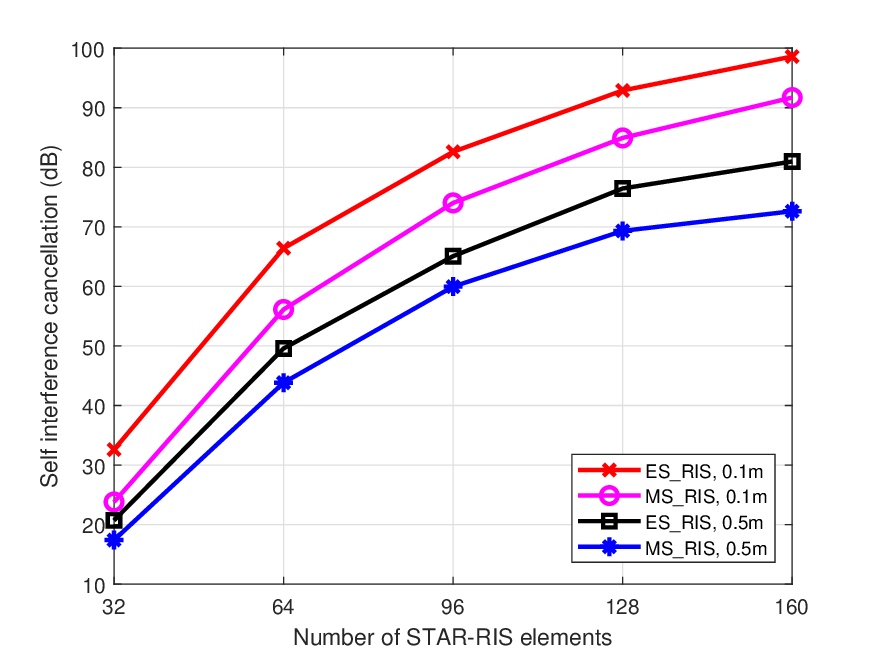}
       \caption{SIC performance for the STAR-RIS empowered FD system with different number of STAR-RIS elements.}
        \label{figR2}
         \vspace{-1em}
\end{figure}

As illustrated in Fig. \ref{figR1}, by integrating STAR-RIS into the FD system, we can maximize the rate transmitted from the FD system's transmit antennas to the receiver (Bob) while meeting fixed self-interference requirements. Since each element of STAR-RIS can either refract or reflect, each STAR-RIS element can help enhance the signal sent to Bob. In this case, we considered the highly flexible ES-RIS and the easily implementable MS-RIS. As shown in Fig. \ref{figR1}, ES-RIS performs best in the proposed FD system because each element can choose to both reflect and refract, offering more control flexibility compared to MS-RIS. However, while ES-RIS brings a larger configuration space, it also introduces a problem of higher hardware complexity. Even when considering practical discrete phase levels in the proposed system, the requirements for practical deployment of ES-RIS are higher than those for MS-RIS. Moreover, we considered the impact of the distance between STAR-RIS and FD antennas. In simulations, we set the distances between STAR-RIS and FD antennas at 0.1m and 0.5m, respectively. The results indicated that if STAR-RIS is too far from the antennas, its effectiveness significantly decreases.

One of the most important performance metrics in a FD system is the effectiveness of SIC. While ensuring a certain reception rate for Bob, we experimented with the impact of varying numbers of STAR-RIS elements on SIC. Simulation results presented in Fig \ref{figR2} demonstrate that as the number of elements in a STAR-RIS increases, the effectiveness of SIC significantly improves. To be Specific, the results indicate that with a substantial number of elements, residual self interference can be reduced to almost negligible levels, while ensuring that the data rate received at the FD receive antennas remains approximately 13 bps/Hz. Therefore, we can conclude that there is a positive correlation between the number of STAR-RIS elements and the performance gains in FD systems. Moreover, in the simulations, we employed the proposed generative deep learning algorithm, which can fast produce prediction results in dynamic and complex environments, offering a low complexity solution for future complex and large-scale wireless communication systems.

The integration of STAR-RIS technology in FD systems marks a significant leap forward in overcoming one of the most important issue in wireless communication technology—self-interference. By using its dual capability to refract and reflect signals, STAR-RIS not only efficiently addresses self interference but also paves the way for more robust, energy-efficient, and high-capacity wireless networks. As this technology continues to develop, its implications will extend beyond the interference cancellation, opening new pathways for advanced signal processing techniques for enhancing the performance of networks with large number of users, and improving the quality of service across the whole wireless communication networks. Looking ahead, the ongoing research and future advancements in STAR-RIS technology hold the promise of revolutionizing communication infrastructures, making it an indispensable component in the evolution of next-generation FD wireless systems. This innovation is a key step towards achieving the goal of ultra reliable, high throughput, low latency, and seamlessly connected global communication networks.

\section{Future Research Directions}\label{sec:Future}
As global communication demands continue to grow and the network environment becomes increasingly complex, traditional communication technologies face multifaceted challenges. FD communication technology is severely limited by the issue of self interference, while the emergence of STAR-RIS provides an innovative solution to this problem. By dynamically adjusting the electromagnetic response of its surface units, STAR-RIS can precisely control the phase and amplitude of reflected signals, effectively eliminating or reducing self interference and enabling high speed data transmission in FD systems. The integration of this technology not only significantly enhances the performance of communication systems but also expands their application range, from high speed data transmission in densely populated urban areas to basic communication services in remote regions. Furthermore, the intelligent features of STAR-RIS enable it to adapt to environmental changes. STAR-RIS can optimize signal transmission paths, and improve the overall reliability and efficiency of the FD system. Therefore, future research directions will include several key areas of wireless communications such as channel estimation, integrated sensing and communication, physical layer security, and space-air-ground networks.

\subsection{Channel Estimation}
In STAR-RIS empowered FD systems, precise channel estimation is the key to achieving efficient communication. As STAR-RIS can control the propagation paths of signals, it enables more accurate channel modeling and estimation in complex signal environments. Future research could develop algorithms that can fast adapt to environmental changes and optimize channel estimation parameters in real-time. For instance, deep learning-based models could be utilized to predict channel characteristics under non-line-of-sight (NLOS) conditions and dynamically adjust the reflective properties of STAR-RIS units to optimize signal quality and system performance. Furthermore, with the advancement of quantum computing and edge computing technologies, future channel estimation methods could integrate these technologies to handle larger data scales, providing faster and more accurate channel estimation results. On this basis, FD systems could effectively eliminate self interference without additional hardware support, significantly enhancing communication reliability and efficiency.

\subsection{Dual-Sided STAR-RIS and Active RIS}
Recently, the dual-sided STAR-RIS and active RIS have started to gain attention in wireless research works. For FD systems, it is possible for the same RIS panel to handle both transmitting and receiving information simultaneously, which is particularly advantageous for dual-sided STAR-RIS. Therefore, integrating dual-sided RIS with FD systems will be a significant direction for future research. Moreover, considering that the passive reflection of RIS sometimes cannot provide sufficiently strong signals, the enhanced capabilities of active RIS can offer better information transmission for future FD systems.

\subsection{Integrated Sensing and Communication}
Integrated sensing and communication play a crucial role in STAR-RIS-empowered FD communication systems. By integrating advanced sensing capabilities, STAR-RIS empowered FD systems not only improves communication efficiency but also optimizes communication strategies by monitoring environmental changes in real-time. For example, STAR-RIS empowered FD systems could use its sensor network on the surface to collect environmental data and adjust its reflective properties according to these data to counteract signal attenuation or interference caused by environmental changes. European Telecommunications Standards Institute (ETSI) has also proposed in its standards ETSI-GR-RIS-002 that RIS should have sensing capabilities to assist with control and optimization needs. Moreover, future research could explore how to integrate STAR-RIS empowered FD systems with mobile platforms such as unmanned aerial vehicles (UAVs) and satellites to create dynamic integrated sensing and communication networks, greatly enhancing capabilities for disaster response and emergency communications.

\subsection{Physical Layer Security}
Physical layer security is an expanding research area in FD communications. STAR-RIS technology offers strong advantages in this area, as it can enhance communication secrecy performance by controlling the propagation paths of signals. Future research could utilize STAR-RIS to reconfig complex signal propagation environments in FD systems to confuse and prevent potential attackers and eavesdroppers from accessing secure information. For example, STAR-RIS could dynamically change the direction and strength of legitimate signal propagation and send jamming signals, making it difficult for eavesdropping devices to capture complete signals or decipher signal content. Thus, ongoing research and development in physical layer security will ensure that the future STAR-RIS empowered FD communication systems can resist more complex and diverse threats in the future.

\subsection{Space-Air-Ground Networks}
The construction of space-air-ground networks is an important trend in the development of future communication networks, and the application of STAR-RIS technology has the potential to play a key role in this area. By deploying STAR-RIS empowered FD systems across different platforms, such as ground stations, UAVs, and satellites, a fully covered and high-speed communication network can be achieved to effectively support seamless connections from urban to remote areas. Future research should focus on how to effectively integrate and coordinate STAR-RIS empowered FD systems across these different platforms to ensure seamless information transfer, optimizing data flow and service quality. Moreover, with the development of next-generation communication technologies, such as beyond-fifth-generation (B5G) and 6G, research on how to utilize STAR-RIS and FD technologies to support higher data rates and lower latency will be important for enhancing space-air-ground networks performance. In summary, the expansion of space-air-ground networks will make FD communication networks more flexible and resilient, capable of handling various complex and extreme environmental conditions, meeting the needs of future societies and technological demands.

\section{Conclusions}\label{sec:con}
This paper primarily highlighted the potential applications of STAR-RIS in FD systems. Due to the challenges associated with traditional FD systems, such as self interference elimination demanding significant hardware resources and energy consumption, we utilized STAR-RIS to assist in SIC within FD systems. By introducing the simultaneous reflection and transmission capabilities of STAR-RIS, its ability to flexibly configure the signal environment and its low-cost characteristics, we proposed utilizing the reflective surface of STAR-RIS to eliminate self interference, while employing its transmissive surface to send the transmit signal in the FD system. Compared to traditional approaches, we adopted deep learning techniques, specifically generative model-based design schemes, to optimize the proposed STAR-RIS empowered FD system. Simulation results demonstrated the effectiveness of the proposed solution and analyzed the performance of STAR-RIS empowered FD systems under various scenarios and configurations. Given the significance of FD systems in wireless communications, the proposed STAR-RIS-based SIC scheme will provide valuable insights for future communication systems. For more complex RIS-assisted FD systems in the future, we estimate using more deep learning algorithms such as reinforcement learning, to guide the system optimization strategy. Moreover, considering the variability of environments in which FD systems are applied, future generative artificial intelligence can better assist in the adaptive optimization of the FD nodes within wireless networks.


\bibliographystyle{ieeetr}
\bibliography{ref}

\begin{thebibliography}{10}

\bibitem{SE_refer1}
W.~Chen, X.~Lin, J.~Lee, A.~Toskala, S.~Sun, C.~F. Chiasserini, and L.~Liu,
  ``{5G}-advanced toward {6G}: Past, present, and future,'' {\em IEEE Journal
  on Selected Areas in Communications}, vol.~41, pp.~1592--1619, Jun. 2023.

\bibitem{FD_survey2}
M.~Mohammadi, Z.~Mobini, D.~Galappaththige, and C.~Tellambura, ``A
  comprehensive survey on full-duplex communication: Current solutions, future
  trends, and open issues,'' {\em IEEE Communications Surveys \& Tutorials},
  vol.~25, no.~4, pp.~2190--2244, Fourthquarter. 2023.

\bibitem{FD_survey_multisce}
K.~E. Kolodziej, B.~T. Perry, and J.~S. Herd, ``In-band full-duplex technology:
  Techniques and systems survey,'' {\em IEEE Transactions on Microwave Theory
  and Techniques}, vol.~67, pp.~3025--3041, Jul. 2019.

\bibitem{SIC_new1}
M.~A. Islam, G.~C. Alexandropoulos, and B.~Smida, ``Joint analog and digital
  transceiver design for wideband full duplex {MIMO} systems,'' {\em IEEE
  Transactions on Wireless Communications}, vol.~21, pp.~9729--9743, Nov. 2022.

\bibitem{RIS_survey}
Y.~Liu, X.~Liu, X.~Mu, T.~Hou, J.~Xu, M.~Di~Renzo, and N.~Al-Dhahir,
  ``Reconfigurable intelligent surfaces: Principles and opportunities,'' {\em
  IEEE Communications Surveys \& Tutorials}, vol.~23, pp.~1546--1577, 3rd
  Quart. 2021.

\bibitem{9475160}
C.~Pan, H.~Ren, K.~Wang, J.~F. Kolb, M.~Elkashlan, M.~Chen, M.~Di~Renzo,
  Y.~Hao, J.~Wang, A.~L. Swindlehurst, X.~You, and L.~Hanzo, ``Reconfigurable
  intelligent surfaces for 6{G} systems: Principles, applications, and research
  directions,'' {\em IEEE Communications Magazine}, vol.~59, no.~6, pp.~14--20,
  Jun. 2021.

\bibitem{STAR-2}
X.~Mu, Y.~Liu, L.~Guo, J.~Lin, and R.~Schober, ``Simultaneously transmitting
  and reflecting {(STAR)} {RIS} aided wireless communications,'' {\em IEEE
  Transactions on Wireless Communications}, vol.~21, no.~5, pp.~3083--3098,
  May. 2022.

\bibitem{STAR-1}
Y.~Liu, X.~Mu, J.~Xu, R.~Schober, Y.~Hao, H.~V. Poor, and L.~Hanzo, ``{STAR}:
  Simultaneous transmission and reflection for 360° coverage by intelligent
  surfaces,'' {\em IEEE Wireless Communications}, vol.~28, pp.~102--109, Dec.
  2021.

\bibitem{10194555}
S.~Fang, G.~Chen, P.~Xiao, K.-K. Wong, and R.~Tafazolli, ``Intelligent omni
  surface-assisted self-interference cancellation for full-duplex {MISO}
  system,'' {\em IEEE Transactions on Wireless Communications}, vol.~23, no.~3,
  pp.~2268--2281, Mar. 2024.

\bibitem{IOS_prototype}
H.~Zhang, S.~Zeng, B.~Di, Y.~Tan, M.~Di~Renzo, M.~Debbah, Z.~Han, H.~V. Poor,
  and L.~Song, ``Intelligent omni-surfaces for full-dimensional wireless
  communications: Principles, technology, and implementation,'' {\em IEEE
  Communications Magazine}, vol.~60, pp.~39--45, Feb. 2022.

\bibitem{FD_survey1}
B.~Smida, A.~Sabharwal, G.~Fodor, G.~C. Alexandropoulos, H.~A. Suraweera, and
  C.-B. Chae, ``Full-duplex wireless for {6G}: Progress brings new
  opportunities and challenges,'' {\em IEEE Journal on Selected Areas in
  Communications}, vol.~41, pp.~2729--2750, Sep. 2023.

\bibitem{SIC_power_3}
M.~Essawy, K.~Rashed, A.~Aghighi, and A.~S. Natarajan, ``A low-noise dual-path
  self-interference cancellation architecture for {W}att-level {TX} power
  handling in simultaneous transmit and receive,'' {\em IEEE Journal of
  Solid-State Circuits}, vol.~59, no.~5, pp.~1337--1350, May. 2024.

\bibitem{7901493}
J.~Zhou, N.~Reiskarimian, J.~Diakonikolas, T.~Dinc, T.~Chen, G.~Zussman, and
  H.~Krishnaswamy, ``Integrated full duplex radios,'' {\em IEEE Communications
  Magazine}, vol.~55, no.~4, pp.~142--151, Apr. 2017.

\bibitem{10466589}
Y.~Wen, G.~Chen, S.~Fang, Z.~Chu, P.~Xiao, and R.~Tafazolli,
  ``{STAR}-{RIS}-assisted-full-duplex jamming design for secure wireless
  communications system,'' {\em IEEE Transactions on Information Forensics and
  Security}, vol.~19, pp.~4331--4343, May. 2024.

\bibitem{9847234}
C.~Huang, G.~Chen, J.~Tang, P.~Xiao, and Z.~Han, ``Machine-learning-empowered
  passive beamforming and routing design for {M}ulti-{RIS}-assisted multihop
  networks,'' {\em IEEE Internet of Things Journal}, vol.~9, no.~24,
  pp.~25673--25684, Dec. 2022.

\end{thebibliography}

\noindent\\
\textbf{Chong Huang} (Member, IEEE) received the Ph.D. degree in wireless communications from University of Surrey in 2022. He is currently a Research Fellow with the Institute for Communication Systems, 5GIC \& 6GIC, University of Surrey. His research interests include deep learning, cooperative networks, physical layer security, cognitive radio, non-orthogonal multiple access, reconfigurable intelligent surfaces (RIS), federated learning, satellite communications and Internet of Things.\\

\noindent
\textbf{Yun Wen} (Graduate Student Member, IEEE) received the B.E. degree in electrical information engineering from Zhejiang University, China, in 2004, and the M.S. degree in informatics from Kyoto University, Japan, in 2009. He is currently pursuing the Ph.D. degree with the Institute for Communication Systems (ICS), University of Surrey, U.K. His research interests include wireless security, full-duplex communications, satellite communications, and reconfigurable intelligent surfaces (RIS).\\

\noindent
\textbf{Long Zhang} received the doctoral degree in Engineering Science from the University of Oxford in 2019. He is currently a research associate professor in Pengcheng Laboratory and was selected into Peacock Program for Overseas High-Level Talents Introduction in Shenzhen. His research interests include wireless communication, free-space optical wireless communication, Lidar and digital signal processing.\\

\noindent
\textbf{Gaojie Chen} (Senior Member, IEEE) received the B.Eng. and B.Ec. degrees in electrical information engineering and international economics and trade from Northwest University, China, in 2006, and the M.Sc. (Hons.) and Ph.D. degrees in electrical and electronic engineering from Loughborough University, Loughborough, U.K., in 2008 and 2012, respectively. He is currently an Assistant Professor with the Institute for Communication Systems, 5GIC \& 6GIC, University of Surrey, U.K., and a Visiting Research Collaborator with the Information and Network Science Lab, University of Oxford.\\

\noindent
\textbf{Zhen Gao} received the B.S. degree in information engineering from Beijing Institute of Technology, Beijing, China, in 2011, and the Ph.D. degree in communication and signal processing from the Tsinghua National Laboratory for Information Science and Technology, Department of Electronic Engineering, Tsinghua University, Beijing, in 2016. He is currently an Assistant Professor with Beijing Institute of Technology. His research interests are in wireless communications, with a focus on multi-carrier modulations, multiple antenna systems, and sparse signal processing.\\

\noindent
\textbf{Pei Xiao} (Senior Member, IEEE) received the Ph.D. degree from the Chalmers University of Technology, Göteborg, Sweden, in 2004. He is a Professor of Wireless Communications with the Institute for Communication Systems, home of 5GIC and 6GIC with the University of Surrey, Guildford, U.K. He is currently the Technical Manager of 5GIC/6GIC, leading the research team in the new physical-layer work area, and coordinating/supervising research activities across all the work areas.

\end{document}